\documentclass[eqsecnum,showpacs,pre,twocolumn]{revtex4}

\usepackage{graphicx}
\usepackage{amsmath}
\usepackage{bm}
\usepackage{epsfig}
\usepackage{amsfonts}
\usepackage{amssymb}%
\begin{document}
\title{Logarithmic Corrections in Directed Percolation}
\author{Hans-Karl Janssen}
\affiliation{Institut f\"{u}r Theoretische Physik III, 
Heinrich-Heine-Universit\"{a}t 40225,
D\"{u}sseldorf, 
Germany}
\author{Olaf Stenull}
\affiliation{Department of Physics and Astronomy, 
University of Pennsylvania, 
Philadelphia, PA 19104, 
USA}
\date{\today}

\begin{abstract}
\noindent We study directed percolation at the upper critical transverse dimension $d=4$, where critical fluctuations induce logarithmic corrections to the leading (mean-field) behavior. Viewing directed percolation as a kinetic process, we address the following properties of directed percolation clusters: the mass (the number of active sites or particles), the radius of gyration and the survival probability. Using renormalized dynamical field theory, we determine the leading and the next to leading logarithmic corrections for these quantities. In addition, we calculate the logarithmic corrections to the equation of state that describes the stationary homogeneous particle density in the presence of a homogeneous particle source.  
\end{abstract}
\maketitle

\section{Introduction}

Directed percolation (DP)~\cite{Hi01} is an anisotropic variant of the usual isotropic percolation (IP)~\cite{reviewsIP} in which an effect or activity can percolate only along a given preferred (longitudinal) direction. DP is perhaps the simplest model leading to self-affine fractals. It has many potential applications, including fluid flow through porous media under gravity, hopping conductivity in a strong electric field~\cite{vanLien_shklovskii_81}, crack propagation~\cite{kertez_vicsek_80}, and the propagation of surfaces at depinning transitions in one dimension~\cite{depinning}. Moreover, it is related to self-organized critical models~\cite{soc}.

Often the longitudinal direction is viewed as time and DP is interpreted as a spreading process. In this dynamic interpretation DP has become famous as the the generic universality class for phase transitions between an active and an absorbing inactive state~\cite{Ja81,Gra82}. The perhaps most intuitive spreading process belonging to the DP universality class is the so called simple epidemic process (SEP). In epidemic processes, individuals (also referred to as particles and for simplicity assumed to be located on the sites of a $d$-dimensional lattice) are either susceptible, infected or immune. At time $t$ an infected particle can randomly infect any of its susceptible neighbors with a certain activation rate. At $t+1$ the newly infected particles are capable of infecting their susceptible neighbors and so on. With a certain deactivation rate any infected particle may become immune. Depending on the difference $\tau$ between the activation and the deactivation rate, the process is endemic or epidemic. For $\tau<\tau_c$ the process dies out after a finite time.  For $\tau>\tau_c$ the process spreads over the entire lattice and approaches a homogenous steady state. The point $\tau=\tau_c$  marks a non-equilibrium phase transition. There are 2 basic variants of epidemic processes. In the SEP there is a chance that an immune particle becomes susceptible again. The spatio-temporal patterns generated by the SEP are DP clusters~\cite{Ja81,Gra82,Ja01}. If immune particles remain immune at all times, one has the so called general epidemic process (GEP). The clusters of immune particles generated by the GEP are IP clusters. 

Since DP represents the generic universality class for phase transitions into absorbing inactive states, it frequently occurs that work on non-equilibrium phase transitions addresses the question, whether a given system belongs to the DP universality class or not. Usually this is done via the fluctuation induced anomalous critical exponents. However, directly at the upper critical dimension, i.e., in $d=4$ transversal dimensions for DP, the leading scaling behavior is purely of mean-field-type and there are no anomalous critical exponents. However, fluctuations lead logarithmic corrections to the mean-field behavior. Just as the anomalous critical exponents, the logarithmic corrections can be used to decide if a given system belongs to the DP universality class. With the computer resources available today, numerical simulation on non-equilibrium systems explore more and more often high spatial dimensions. Simulations with reliable statistics of such systems in 4 dimensions are within reach today. Hence, we feel that it becomes important to know logarithmic corrections for the DP universality class.

The leading logarithmic corrections are fairly easy to extract from the known renormalization group (RG) results on DP. Astonishingly, this has not been done to date, at least to our knowledge. It has to be expected, though, that knowing the leading logarithmic corrections is not sufficient to obtain a decent agreement between theory and simulations. This expectation is based on the experiences that have been made for another system for which logarithmic correction have been studied intensively by numerical and analytical means, viz.\ linear polymers~\cite{GHS94,GHS99}. Numerical work on DP in $d=4$ in progress seems to corroborate our expectation~\cite{Lue03}.

The aim of this paper is to derive analytically logarithmic correction for DP up to and including the next to leading order. We focus three dynamical observables that are well suited for investigation by numerical simulations, namely the number $N(t)$ of infected particles at time $t$ generated by a seed at the spatio-temporal origin $(\mathbf{x}=\mathbf{0},t=0)$, their mean distance $R(t)$ from the origin (radius of gyration) as well as he survival probability $P(t)$ of the corresponding cluster. Furthermore, we determine logarithmic corrections for the DP equation of state (EQS) that relates the homogeneous particle density $M$ of the stationary state to $\tau$ and an auxiliary constant homogeneous particle source $h$. For dimensions below 4, this EQS is known to 2-loop order~\cite{JaKuOe99}. Its logarithmic correction have not been addressed hitherto.

Of course, logarithmic corrections to dynamic quantities like $N(t)$ are not only relevant for DP; they are likewise important for dynamic IP at the respective upper critical dimension 6. For logarithmic corrections to dynamic IP we refer to Ref.~\cite{JaSt_u1}. Also, logarithmic correction influence the static properties of IP clusters, like for example their various fractal dimensions, and their transport properties. For logarithmic corrections in static IP, see Ref.~\cite{StJa03}.

The outline of our paper is as follows. In Sec.~\ref{reviewDP} we briefly review the SEP as a dynamic model for DP.  We sketch the renormalized field theory of the SEP and cite previous RG results. Furthermore, we  explain by solving the RG equation in $d=4$ how logarithmic corrections arise in DP. In Section~\ref{logCorr} we derive the logarithmic corrections for the aforementioned dynamic observables. Section~\ref{EQS} treats logarithmic corrections of the mean-field equation of state. In Sec.~\ref{concludingRemarks} we give a few concluding remarks. Details of our diagrammatic perturbation calculation are relegated to an Appendix.

\section{A brief review of directed percolation and its dynamical field theory}
\label{reviewDP} 

This section is intended to provide the reader with background on the dynamical field theory of DP and to establish notation. Moreover, it demonstrates how logarithmic correction emerge in the RG framework by solving the RG equation directly in d=4. 

\subsection{Modelling directed percolation}
There are basically two complimentary approaches to model DP. The first approach is based on bond percolation and assigns a direction to the bonds. An example for this kind of model is the random resistor diode network, see, e.g., Refs.~\cite{perc,redner_mult,janssen_stenull_prerapid_2001}. In the other approach one models DP as a kinetic growth process~ \cite{Mo77,Ba85,Mu89}, viz.\ the SEP that we elaborated on in the introduction. Here, we will take the latter route. 

On mesoscopic scales it makes sense to describe the SEP in terms of the density $n(\mathbf{x},t)$ of infected particles at time $t$ and space coordinate $\mathbf{x}$. It is well known that the Langevin equation (in the Ito sense) governing the time evolution of this density is given by~ \cite{Ja81}
\begin{subequations}
\label{langAndNoise}%
\begin{align}
\lambda^{-1}\partial_{t}n(\mathbf{x},t)  &  =\nabla^{2}n(\mathbf{x},t)-\tau
n(\mathbf{x},t)-\frac{g}{2}n(\mathbf{x},t)^{2}
\nonumber \\
&+\zeta(\mathbf{x},t)\,,\label{LangEq}\\
\overline{\zeta(\mathbf{x},t)\zeta(\mathbf{x}^{\prime},t^{\prime})}  &
=\lambda^{-1}g^{\prime}n(\mathbf{x},t)\delta(t-t^{\prime})\delta
(\mathbf{x}-\mathbf{x}^{\prime})\,. \label{Noise}%
\end{align}
\end{subequations}
The parameter $\tau$ is essentially the rate
difference mentioned in the introduction and hence specifies the deviation
from criticality. $\lambda$ represents a kinetic coefficient. $\zeta(\mathbf{x},t)$ is a Gaussian
random field that subsumes reaction noise and otherwise
neglected microscopic details.  $\overline{\cdots}$ stands for averaging over the
distribution of the noise. The right hand side of Eq.~(\ref{Noise}) goes to zero for vanishing $n(\mathbf{x},t)$ to enable the existence of the absorbing state.
Contributions to Eqs.~(\ref{langAndNoise}) that are of higher order in the field or the derivatives turn out to be irrelevant in the sense of the RG. For example, a diffusional noise
contribution can be neglected.

Langevin equations are fairly intuitive and thus provide a natural starting point in the mesoscopic description of stochastic processes. Prevalent alternative forms of mesoscopic description are Fokker-Planck equations as well as dynamic functionals~\cite{Ja76,DeDo76,Ja92}. Dynamic functionals, also known as response functionals, are best suited for the application of field theory and RG ideas. This is the form of description that we will use here. The dynamic functional $\mathcal{J}$ for DP has been known for a long time~\cite{Ja81,Ja01}. After exploiting a rescaling form-invariance that allows to equate $g$ and $g^\prime$, $\mathcal{J}$ can be written as
\begin{equation}
\mathcal{J}=\int d^{d}x\,dt\,\lambda\tilde{s}\Big(\lambda^{-1}\partial
_{t}+(\tau-\nabla^{2})+\frac{g}{2}(s-\tilde{s})\Big)s\,. \label{Funkt}%
\end{equation}
The order parameter field $s(\mathbf{x},t)$ is proportional to particle density $n(\mathbf{x},t)$.
$\tilde{s}(\mathbf{x},t)$ is the response field corresponding to $s(\mathbf{x},t)$. Time reflection, also known as duality transformation,
\begin{equation}
\tilde{s}(\mathbf{x},t)\longleftrightarrow-s(\mathbf{x},-t) \, , \label{Zeitsp.}%
\end{equation}
is a symmetry transformation of the dynamic functional~\cite{Ja81}. This, however, is merely an asymptotic symmetry that holds provided that irrelevant terms are absent. When applying RG ideas to calculate leading scaling scaling properties or logarithmic corrections, one has to make sure that this symmetry is preserved. On the other hand, if one is interested in corrections to scaling stemming from irrelevant contributions to the functional~(\ref{Funkt}), then one has to admit composite fields that break the symmetry~(\ref{Zeitsp.}). 

\subsection{Renormalization and scaling}

The great virtue of response functional $\mathcal{J}$ is that it allows for systematic perturbation calculation in the coupling constant $g$ that resembles many features of, and allows to glean techniques from, the well established diagrammatic perturbation treatments of equilibrium critical phenomena. The most economic way to actually do these calculations is to use dimensional regularization
in conjunction with minimal subtraction (minimal renormalization). For background on these methods we refer to Refs.~\cite{Am84,ZJ96}. An appropriate renormalization scheme is
\begin{subequations}
\label{RenSch}%
\begin{align}
s\rightarrow\mathring{s}=  &  Z^{1/2}s\,,\quad\tilde{s}\rightarrow
\mathring{\tilde{s}}=Z^{1/2}\tilde{s}\,,\\
\lambda\rightarrow\mathring{\lambda}=  &  Z^{-1}Z_{\lambda}\lambda\,,\quad
\tau\rightarrow\mathring{\tau}=Z_{\lambda}^{-1}Z_{\tau}\tau\,,\\
g\rightarrow\mathring{g}=  &  Z_{\lambda}^{-1}Z^{-1/2}Z_{u}^{1/2} G_{\varepsilon}^{-1/2} \mu^{\varepsilon/2 } \, u^{1/2} \, ,
\end{align}
\end{subequations}
where the symbol $\mathring{}$\thinspace\ indicates unrenormalized quantities. The factor $\mu^{\varepsilon/2}$, where $\mu$ is an arbitrary external inverse length scale and $\varepsilon=4-d$ measures the deviation from the upper critical dimension makes the renormalized coupling constant $u$ dimensionless.  The quantity $G_{\varepsilon}=\Gamma(1+\varepsilon/2)/(4\pi)^{d/2}$ naturally appears in the computation of Feynman diagrams and is included here for later convenience. In minimal renormalization the critical point value $\tau=\tau_{c}$ is formally set to zero by the perturbation expansion. In general, however, $\tau_{c}$ is a non-analytical function of the coupling constant. Thus, an implicit additive renormalization $\tau-\tau_{c}\rightarrow\tau$ is concealed in the minimal renormalization procedure. The renormalization factors $Z$, $Z_{\lambda}$, $Z_{\tau}$, and $Z_{u}$ are known to 2-loop order~\cite{Ja81,Ja01}. 

The critical behavior of the Green's functions $G_{n,\tilde{n}}=\langle\lbrack
s]^{n}[\tilde{s}]^{\tilde{n}}\rangle^{(\text{cum})}$, where $\langle \cdots \rangle^{(\text{cum})}$ denotes the cumulants with respect to the statistical weight $\exp (-\mathcal{J})$, is governed by the
Gell-Mann--Low renormalization group equation (RGE)
\begin{equation}
\Big[\mathcal{D}_{\mu}+\frac{n+\tilde{n}}{2}\gamma\Big]G_{n,\tilde{n}%
}(\{\mathbf{r},t\};\tau,u;\lambda,\mu)=0  \label{RGG}%
\end{equation}
with the RG differential operator $\mathcal{D}_{\mu}$ given by
\begin{equation}
\mathcal{D}_{\mu}=\mu\partial_{\mu}+\lambda\zeta\partial_{\lambda}+\tau
\kappa\partial_{\tau}+\beta\partial_{u}\,. \label{RGOp}%
\end{equation}
The Wilson functions featured in the RGE are known to 2-loop order~\cite{Ja81,Ja01},
\begin{subequations}
\label{RGFunkt}%
\begin{align}
\gamma &  =-\frac{u}{4}+\Big(6-9\ln\frac{4}{3}\Big)\frac{9u^{2}}{32}%
+O\left( u^3  \right)\,,\\
\zeta &  =-\frac{u}{8}+\,\Big(17-2\ln\frac{4}{3}\Big)\frac{u^{2}}{256}%
+O\left( u^3  \right) \,,\\
\kappa &  =\frac{3u}{8}-\,\Big(7+10\ln\frac{4}{3}\Big)\frac{7u^{2}}%
{256}+O\left( u^3  \right) \,,\\
\beta &  =-\varepsilon u+\frac{3u^{2}}{2}-\Big(169+106\ln\frac{4}{3}%
\Big)\frac{u^{3}}{128}+O\left( u^4  \right) \,.
\end{align}
\end{subequations}

The RGE can be solved by the method of characteristics. The idea behind this method is to consider all the scaling parameters as a function of a single flow parameter $l$. One sets sets up characteristic equations that describe how the scaling parameters transform under a change of $l$. The characteristic for the momentum scale $\mu$ is particularly simple and has the solution $\bar{\mu}(l)=\mu l$, i.e., a change of $l$ corresponds to a change of the external momentum scale. With help of the solution to the remaining characteristics one obtains 
\begin{align}
\label{GrFuSkal}
&  G_{n,\tilde{n}}(\{\mathbf{x},t\};\tau,u;\lambda,\mu)  =\,\bigl[(\mu l)^{d}Z(l)\bigr]^{(n+\tilde{n})/2}\\
&  \times\,G_{n,\tilde{n}}(\{l\mu\mathbf{x},Z_{\lambda}(l)(l\mu)^{2}\lambda
t\};Z_{\tau}(l)\tau/(\mu l)^{2},\bar{u}(l);1,1)\,.\nonumber
\end{align}
as a solution to the RGE. At this stage the scaling solution~(\ref{GrFuSkal}) is still rather formal since $Z(l)$, $Z_{\lambda}(l)$, $Z_{\tau}(l)$ and $\bar{u}(l)$ require specification. Below the upper critical dimension, these quantities display power law behavior described by the well known critical exponents of the DP universality class. Directly in $d=4$, their depend logarithmically on $l$ and hence their behavior is qualitatively different from the lower dimensional case.

\subsection{General form of the logarithmic corrections}
\label{genFormLog}

Now we will state and solve the characteristics directly for $d=4$. The Wilson functions cited in Eqs.~(\ref{RGFunkt}) are central ingredients of the characteristics. To make the notion more economic, we will write $f(u)=f_{0}+f_{1}u+f_{2}u^{2}+\cdots$ with $f$ standing ambiguously for $\gamma$, $\zeta$, $\kappa$, and $\beta$. We expect the meaning of the coefficients $f_{0}$, $f_{1}$ and so on to be evident.

The characteristic for the dimensionless coupling constant $u$ is given by
\begin{equation}
l\frac{dw}{dl}=\beta(w)\, \label{Char-u}%
\end{equation}
where we abbreviated $w=\bar{u}(l)$. Solving this differential equation for
$\varepsilon=0$ yields
\begin{equation}
l=l(w)=l_{0}w^{-\beta_{3}/\beta_{2}^{2}}\exp\bigg[-\frac{1}{\beta_{2}%
w}+O(w)\bigg]\,, \label{l(w)}%
\end{equation}
where $l_{0}$ is an integration constant. The remaining characteristics are
all of the same structure, namely
\begin{equation}
l\frac{d\ln Q(w)}{dl}=q(w)\,.
\end{equation}
Here, $Q$ is a placeholder for $Z$, $Z_{\tau}$, and $Z_{\lambda}$,
respectively. $q$ is a wildcard for $\gamma$, $\kappa$, and $\zeta$,
respectively. Exploiting $ld/dl=\beta d/dw$ we obtain the solution
\begin{equation}
Q(w)=Q_{0}w^{q_{1}/\beta_{2}}\exp\bigg[\frac{(q_{2}\beta_{2}-q_{1}\beta_{3}%
)}{\beta_{2}^{2}}w+O(w^{2})\bigg]\,, \label{Q(w)}%
\end{equation}
where $Q_{0}$ symbolizes a non universal integration constant.

The flow parameter introduced via the characteristics is arbitrary. This arbitrariness has an important virtue, viz. $l$ can be chosen so that one of the relevant variables $\mathbf{x}$, $t$, or $\tau^{-1}$ effectively acquires a finite value in the scaling limit. In this paper we are interested in time-dependent quantities and hence we choose
\begin{equation}
Z_{\lambda}(w)(l\mu)^{2}\lambda t=X_{0}\,, \label{Wahl_X}%
\end{equation}
where $X_{0}$ is a constant of order unity. With this choice $w$ and $l$ tend
to zero for $\lambda\mu^{2}t\rightarrow\infty$. Instead of using the original $t$, we find it convenient to use 
\begin{equation}
s=\frac{\beta_{2}}{2}\ln\bigl(t/t_{0}\bigr)=\frac{3}{4}\ln\bigl(t/t_{0}%
\bigr)\,, \label{s(t)}%
\end{equation}
as our time variable. Here, $t_{0}$ is a non universal time constant proportional to $X_{0}$. From Eq.~(\ref{l(w)}) and Eq.~(\ref{Q(w)}), specialized to $Z_{\lambda}$, we obtain
\begin{equation}
s=w^{-1}-a_{0}\ln w+O(w) \, . \label{t(w)}
\end{equation}
The constant $a_{0}$ is given by
\begin{equation}
a_{0}=\frac{\beta_{2}\zeta_{1}-2\beta_{3}}{2\beta_{2}}=\frac{157}{192}%
+\frac{53}{96}\ln\frac{4}{3}=0.976533\,.
\end{equation}
Finally, we find by using Eq.~(\ref{t(w)}) 
\begin{equation}
w=s^{-1}\exp\bigg[a_{0}\frac{\ln s}{s}+O\Big(\frac{\ln^{2}s}{s^{2}},\frac{\ln
s}{s^{2}},\frac{1}{s^{2}}\Big)\bigg] \label{w(s)}%
\end{equation}
for the dimensionless coupling constant as a function of time.

\section{Logarithmic corrections for the number of active particles etc.}

\label{logCorr} 

\subsection{Number of active particles}
\label{partNum}

At criticality ($\tau=0$), the number of active particles generated by a seed at the origin is related to the Green's function $G_{1,1}$ via
\begin{equation}
N(t)  =\int d^{d}x\, \, G_{1,1}(\mathbf{x},t;0,u;\lambda,\mu)
\end{equation}
Using the general scaling result~(\ref{GrFuSkal}) we can express the scaling behavior of $N(t)$ as
\begin{align}
\label{N(t)1}
N(t) &  =Z(w)\int d^{d}x\,(\mu l)^{d}\nonumber\\
&  \times G_{1,1}(l\mu\mathbf{r},Z_{\lambda}(w)(l\mu)^{2}\lambda
t;0,w;1,1)
\nonumber \\
&  =Z(w) \, G_{1,1}(\mathbf{q}=0,X_{0};0,w;1,1)
\end{align}
with $Z(w)$ given by Eq.~(\ref{Q(w)}) specialized to $Q=Z$. 

\begin{figure}[ptb]
\includegraphics[width=2.25cm]{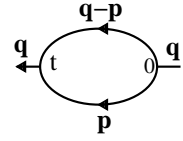}\caption{Self-energy $\Sigma
(\mathbf{q},t)$ at 1-loop order.}%
\label{selfEn}%
\end{figure}

\begin{figure}[ptb]
\includegraphics[width=8.4cm]{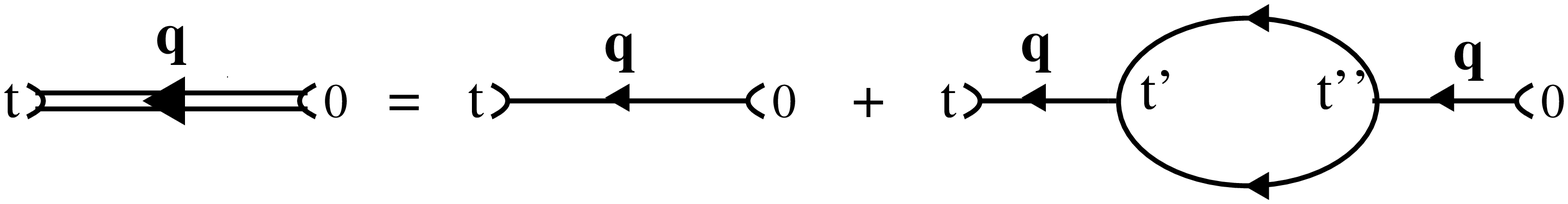}\caption{Dyson equation~(\ref{Dys-Gl}) to
1-loop order.}%
\label{dysonGl}%
\end{figure}

Note that the Green's function in the last line of Eq.~(\ref{N(t)1}) depends on the renormalized coupling constant $w$. If we were interested only in the leading logarithmic correction to $N(t)$ we could ignore this subtlety. The higher logarithmic corrections, however, are influenced by the specifics of the Green's function. For the second logarithmic correction, we have to calculate $G_{1,1} (\mathbf{x},t)$ to 1-loop order. The diagrammatic elements required in this calculation are the Gaussian propagator
\begin{equation}
G(\mathbf{q},t)=\theta(t)\,\exp\big[-\lambda(\tau+q^{2})t\big] \, ,\label{Prop}
\end{equation}
where $\theta$ stands for the step function, and the two 3-leg vertices $\lambda g$ and $-\lambda g$. In contrast to the calculation of for example critical exponents, it is not sufficient for our current purposes to consider Feynman diagrams with amputated external legs because $G_{1,1}$ is determined by the Dyson equation
\begin{align}
G_{1,1}(\mathbf{q},t)  &  =G(\mathbf{q},t)+\int_{0}^{t}dt^{\prime}\int
_{0}^{t^{\prime}}dt^{\prime\prime}\,G(\mathbf{q},t-t^{\prime})\nonumber\\
&  \times\Sigma(\mathbf{q},t^{\prime}-t^{\prime\prime})\,G(\mathbf{q}%
,t^{\prime\prime})+\cdots\,. \label{Dys-Gl}%
\end{align}
Here, $\Sigma(\mathbf{q},t)$ stands for the self-energy that is given to 1-loop order by 
\begin{equation}
\Sigma(\mathbf{q},t)=-\frac{\lambda^{2}g^{2}}{2}\int_{\mathbf{p}}%
G(\mathbf{p},t) \, G(\mathbf{q-p},t)\,, \label{Selbst}%
\end{equation}
where $\int_{\mathbf{p}}$ stands for $(2\pi)^{-d/2} \int d^d p$. Diagrammatic representations of the self-energy and the Dyson equation can be found in Fig.~\ref{selfEn} and Fig.~\ref{dysonGl}, respectively.
%%%%%%%%%%%%%%%%%%%%%%%
%%%%%%%%%%%%%%%%%%%%%%%
%%%%%%%%%%%%%%%%%%%%%%%
%%%%%%%%%%%%%%%%%%%%%%%
For details of our calculation we refer to Appendix~\ref{app:G11}. Eventually we find
\begin{equation}
G_{1,1}(\mathbf{q}=0,X_{0};0,w;1,1)=1+A_{N}(X_{0})w+O(w^{2})\,,  \label{A_N}%
\end{equation}
where $A_{N}(X_{0})$ is an amplitude given by
\begin{equation}
A_{N}(X_{0})=\frac{1}{8}\bigl(\mathcal{Z}+1\bigr) \, . \label{A_N(X)}%
\end{equation}
Here, we used the shorthand notation $\mathcal{Z}=\ln(2X_{0})+C_{E}$ with
$C_{E}$ being Euler's constant. 

Now we know all the ingredients that contribute to the leading and the next to leading logarithmic correction to $N(t)$. Collecting from Eqs.~(\ref{N(t)1}), (\ref{Q(w)}) and
(\ref{A_N}) we find
\begin{align}
N(t)  &  =N_{0}\bigl(w^{-1}+B_{N}\bigr)^{1/6}\exp\bigl(-c_{N}w+O(w^{2}%
)\bigr)\nonumber\\
&  =N_{0}^{\prime}\bigl(s+B_{N}\bigr)^{1/6}\Big[1-\frac{b_{N}\ln s+c_{N}}%
{s}\nonumber\\
&  +O\Big(\frac{\ln^{2}s}{s^{2}},\frac{\ln s}{s^{2}},\frac{1}{s^{2}%
}\Big)\Big] \label{Npara1}%
\end{align}
where $N_{0}$ is a non universal constant, $N_{0}^{\prime}$ is a non universal
constant slightly different from $N_{0}$, and $B_{N}=6A_{N}$. The first row of
(\ref{Npara1}) and the result (\ref{t(w)}) constitute a parametric
representation of the tuple $(N,s)$ that is suitable for comparison to
numerical simulations. The parametric representation has the advantage that it
represents a nicely systematic expansion in terms of the coupling constant
$w$. The second row of (\ref{Npara1}) shows the more traditional form. The
constants $b_{N}$, $c_{N}$, and $B_{N}$ are given by
\begin{subequations}
\begin{align}
b_{N}  &  =\frac{a_{0}}{6}=\frac{157}{1152}+\frac{53}{576}\ln\frac{4}%
{3}=0.162755\,,\\
c_{N}  &  =\beta_{3}\frac{\gamma_{1}}{\beta_{2}^{2}}-\frac{\gamma_{2}}%
{\beta_{2}} =\frac{25}{1152}+\frac{161}{576}\ln\frac{4}{3} \nonumber\\
&  =0.102113\,,\\
B_{N}  &  =\frac{3}{4}\bigl(\mathcal{Z}+1\bigr)\,.
\end{align}
Equations~(\ref{s(t)}) and (\ref{Npara1}) show directly that we may eliminate the arbitrary
constant $\mathcal{Z}$ by a rescaling of the non-universal time constant $t_{0}$. However, we keep this constant in our formulas because it can mimic higher, neglected powers in our expansions, i.e., it represents a further constant that can be fitted to simulation results.

\subsection{Radius of gyration}

The mean square distance of the active particles from the origin, also known as their radius of gyration, is defined by
\end{subequations}
\begin{equation}
R(t)^{2}=\frac{\int d^{d}x\,\mathbf{x}^{2}\,G_{1,1}(\mathbf{x},t)}{2d\int
d^{d}x\,G_{1,1}(\mathbf{x},t)}=-\left.  \frac{\partial\ln G_{1,1}%
(\mathbf{q},t)}{\partial q^{2}}\right\vert _{\mathbf{q}=\mathbf{0}}. \label{R-def}%
\end{equation}
The general scaling solution~(\ref{GrFuSkal}) implies that 
\begin{align}
\label{RscaleRel}
R(t)^{2}  =& - \left.  \frac{\partial\ln G_{1,1}((l\mu)^{-1}\mathbf{q},Z_{\lambda
}(w)(l\mu)^{2}\lambda t;0,w;1,1)}{\partial q^{2}}\right\vert _{\mathbf{q}%
=\mathbf{0}}\nonumber\\
   =& - (l\mu)^{-2}\left.  \frac{\partial\ln G_{1,1}(\mathbf{q},X_{0}%
;0,w;1,1)}{\partial q^{2}}\right\vert _{\mathbf{q}=\mathbf{0}}
\end{align}
for $\tau=0$. Equation~(\ref{RscaleRel}) shows that the radius of gyration is, like $N(t)$, affected by the dependence of $G_{1,1}$ on the renormalized coupling constant $w$. Here, however, we need the part of $G_{1,1}$ that is quadratic in the momentum. In Appendix~\ref{app:G11} we calculate that
\begin{align}
\label{combi1}
&  -\left.  \frac{\partial}{\partial q^{2}}\ln G_{1,1}(\mathbf{q}%
,X_{0};0,w;1,1)\right\vert _{\mathbf{q}=\mathbf{0}}\nonumber\\
&  =X_{0}\bigl(1+A_{R}(X_{0})w+O(w^{2})\bigr)\,.
\end{align}
The amplitude that appears here is given by
\begin{equation}
A_{R}(X_{0})=\frac{1}{16}\bigl(\mathcal{Z}-1\bigr)\,, \label{A_R(X)}%
\end{equation}
Combining Eqs.~(\ref{RscaleRel}), (\ref{combi1}) and (\ref{A_R(X)}) as well as the solutions to the appropriate characteristics we obtain
\begin{align}
t^{-1}R^{2}  &  =R_{0}^{2}\bigl(w^{-1}+B_{R}\bigr)^{1/12}\exp\bigl(-c_{R}%
w+O(w^{2})\bigr)\nonumber\\
&  =R_{0}^{\prime2}\bigl(s+B_{R}\bigr)^{1/12}\Big[1-\frac{b_{R}\ln s+c_{R}}%
{s}\nonumber\\
&  +O\Big(\frac{\ln^{2}s}{s^{2}},\frac{\ln s}{s^{2}},\frac{1}{s^{2}%
}\Big)\Big]\,.
\end{align}
with $R_{0}^{2}$ and $R_{0}^{\prime2}$ being non-universal amplitudes.  $b_{R}$, $c_{R}$, and $B_{R}=12A_{R}$ are constants that have the values
\begin{subequations}
\begin{align}
b_{R}  &  =\frac{a_{0}}{12}=\frac{157}{2304}+\frac{53}{1152}\ln\frac{4}%
{3}=0.0813777\,,\\
c_{R}  &  =\frac{\zeta_{1}\beta_{3}}{\beta_{2}^{2}}-\frac{\zeta_{2}}{\beta
_{2}} =\frac{67}{2304}+\frac{59}{1152}\ln\frac{4}{3} \nonumber\\
&  =0.0438136\,,\\
B_{R}  &  =\frac{3}{4}\bigl(\mathcal{Z}-1\bigr)\,.
\end{align}
\end{subequations}

\subsection{Survival probability}

Recently it was demonstrated that the survival probability $P(t)$ of an active cluster
emanating from a seed at the origin can be expressed in terms of the order parameter and the response field as~\cite{Ja02}
\begin{align}
P(t)  &  = -\lim_{k\rightarrow\infty}\langle\mathrm{e}^{-k\mathcal{N}}\tilde
{s}(-t)\rangle \, ,
\end{align} where $\mathcal{N}=\int d^{d}x\,s(\mathbf{x},0)$. This formula is fundamental in that it relates the survival probability unambiguously to the fields inherent in the dynamic functional $\mathcal{J}$. In actual calculations, however, the term $\exp (-k\mathcal{N})$ has to be incorporated into the dynamic functional and one is led to
\begin{eqnarray}
\mathcal{J}_k = \mathcal{J} + \int dt\, k(t)\mathcal{N}(t) 
\end{eqnarray}
instead of the original $\mathcal{J}$. Here, $k(t)=k\delta(t)$ is a source conjugate to the field $s$. Having introduced $\mathcal{J}_k$, one can write
\begin{eqnarray}
\label{newSurvProb}
P(t) &=& -\lim_{k\rightarrow\infty}\langle\tilde
{s}(-t)\rangle_k
\nonumber \\
&=&-G_{0,1}(\mathbf{-}t,\tau,k=\infty,u;\lambda,\mu) \, ,
\end{eqnarray}
where $\langle \cdots \rangle_k$ denotes averaging with respect to  $\mathcal{J}_k$. Note that the explicit term $\exp (-k\mathcal{N})$ is gone. 

At this point we find it worthwhile to annotate an interesting implication of the time reflection symmetry~(\ref{Zeitsp.}). Due to this symmetry the survival probability $P(t)$ is identical to the mean particle density $\rho (t) = \langle s(t)\rangle$ of the dual process starting with a fully occupied initial state, $\rho (0) = \infty$.

To avoid tadpoles in our perturbation
calculation, we carry out the shift $\tilde{s}\rightarrow\tilde{s}+\tilde{M}$ so
that $\langle\tilde{s}\rangle=0$ is restored. $G_{0,1}$ is then nothing but
$\tilde{M}$. The entire procedure leads to the new response functional
\begin{align}
\mathcal{J}_k &=\int d^{d}x\,dt\,\biggl\{ \lambda
\tilde{s}\, \Big(\lambda^{-1}\frac{\partial}{\partial t} + (\tau-g\tilde{M}-\nabla^{2})\nonumber\\
& +\frac{g}{2}(s-\tilde{s})\Big)s  +\frac{\lambda g}{2}\tilde{M}s^{2}
\nonumber \\
& +\Big(-\partial_{t}\tilde{M}+\lambda
\tau\tilde{M}-\frac{\lambda g}{2}\tilde{M}^{2}+k\Big)s\biggr\}\,. \label{J_k}
\end{align}

\begin{figure}[ptb]
\includegraphics[width=3.0cm]{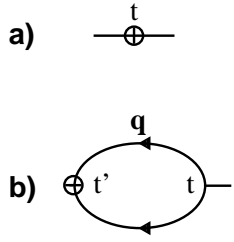}\caption{(a) The new vertex
$-\lambda^{2}g\tilde{M}(t)$ and (b) the 1-loop tadpole diagram $\tilde{T}%
(t)$.}%
\label{tadpole}%
\end{figure}

The diagrammatic elements implicit in $\mathcal{J}_k$ comprise
the two vertices encountered in Sec.~\ref{partNum}. In addition, there is a third
vertex, viz.\ $-\lambda^{2}g\tilde{M}(t)$ as depicted in Fig.~\ref{tadpole}a. The Gaussian propagator
for the new functional has to be determined from the differential equation
\begin{equation}
\Big[\lambda^{-1}\partial_{t}+\tau-g\tilde{M}(t)+q^{2}\Big]\bar{G}%
(\mathbf{q},t,t^{\prime})=\lambda^{-1}\delta(t-t^{\prime})\,. \label{Prop-k}%
\end{equation}
To avoid tadpoles, $\tilde{M}(t)$ must satisfy the differential equation
\begin{equation}
\partial_{t}\tilde{M}(t)-\lambda\tau\tilde{M}(t)+\frac{\lambda g}{2}\tilde
{M}(t)^{2}-k(t)+\tilde{T}(t)=0\,. \label{Dgl-shift}%
\end{equation}
At 1-loop order, the tadpole $\tilde{T}(t)$ is given by the diagram shown in Fig.~\ref{tadpole}b.
%%%%%%%%%%%%%%%%%%%%%%%
%%%%%%%%%%%%%%%%%%%%%%%
Upon solving the differential equations~(\ref{Prop-k}) and (\ref{Dgl-shift}) at the mean-field level we find that the modified Gaussian propagator reads (for details see Appendix~\ref{app:G01})
\begin{equation}
\bar{G}_{0}(\mathbf{q},t,t^{\prime})=\theta(t-t^{\prime})\biggl(\frac
{K_{0}(-t)}{K_{0}(-t^{\prime})}\biggr)^{2}\,\exp\Big[\lambda(\tau
-q^{2})(t-t^{\prime})\Big]\,,  \label{Gmod_0}%
\end{equation}
where 
\begin{equation}
K_{0}(t)=\frac{g}{2\tau}\Big(\mathrm{e}^{\lambda\tau t}-1\Big)\,. \label{K_0}%
\end{equation}
Having the modified Gaussian propagator at our disposal we are in the
position to calculate the diagram depicted in Fig.~\ref{tadpole}b. This calculation leads eventually to 
\begin{align}
&  G_{0,1}(\mathbf{-}X_{0},0,k=\infty,w;1,1)\nonumber\\
&  \propto\,w^{-1/2}\bigl(1+A_{P}(X_{0})w+O(w^{2})\bigr)\,. \label{A_P}%
\end{align}
with the amplitude $A_{P}(X_{0})$ reading
\begin{equation}
A_{P}(X_{0})=\frac{3}{8}\Big(\mathcal{Z}-\frac{3}{2}\Big)\,. \label{A_P(X)}%
\end{equation}
For details on this calculation we refer to Appendix~\ref{app:G01}.

Little further work is required to extract the logarithmic corrections to the survival probability. Recalling the scaling form~(\ref{GrFuSkal}) and our choice for the flow parameter we deduce that, for
$\tau=0$,
\begin{equation}
P(t)=-Z(w)^{1/2}(\mu l)^{2}G_{0,1}(\mathbf{-}X_{0},0,\infty,w;1,1)\,.
\end{equation}
Inserting our results for the characteristics as well as Eq.~(\ref{A_P}) combined with (\ref{A_P(X)}) we obtain
\begin{align}
tP(t)  &  =P_{0}\bigl(w^{-1}+B_{P}\bigr)^{1/2}\exp\bigl(-c_{P}w+O(w^{2}%
)\bigr)\nonumber\label{Ppara}\\
&  =P_{0}^{\prime}\bigl(s+B_{P}\bigr)^{1/2}\Big[1-\frac{b_{P}\ln s+c_{P}}%
{s}\nonumber\\
&  +O\Big(\frac{\ln^{2}s}{s^{2}},\frac{\ln s}{s^{2}},\frac{1}{s^{2}%
}\Big)\Big]\,.
\end{align}
$P_{0}$ and $P_{0}^{\prime}$ are non-universal factors. The
constants $b_{P}$, $c_{P}$, and $B_{P}=2A_{P}$ are given by
\begin{subequations}
\begin{align}
b_{P}  &  =\frac{a_{0}}{2}=\frac{157}{384}+\frac{53}{192}\ln\frac{4}%
{3}=0.488266\,,\\
c_{P}  &  =\frac{2\zeta_{2}-\gamma_{2}}{2\beta_{2}}+\beta_{3}\frac{\gamma
_{1}-2\zeta_{1}}{2\beta_{2}^{2}} =-\frac{7}{384}+\frac{17}{192}\ln\frac{4}{3} \nonumber\\
&  =0.00724268\,,\\
B_{P}  &  =\frac{3}{4}\Big(\mathcal{Z}-\frac{3}{2}\Big)\,.
\end{align}
The constant $\mathcal{Z}$ might be eliminated by the same rescaling of the
non-universal time scale $t_{0}$ as discussed above.

\section{Logarithmic corrections to the equation of state}
\label{EQS}

\subsection{General considerations}

It is well known (see e.g.~\cite{ZJ96}) that the generating functional
$\mathcal{W}\bigl[\tilde{J},J\bigr]$ of the Green's functions
\end{subequations}
\begin{equation}
G_{n,\tilde{n}}(\{\mathbf{x},t\},\{\mathbf{\tilde{x}},\tilde{t}\})=\left.
\frac{\delta^{n+\tilde{n}}\mathcal{W}\bigl[\tilde{J},J\bigr]}{\{\delta
J(\mathbf{x},t)\}^{n}\{\delta\tilde{J}(\mathbf{\tilde{x}},\tilde{t}%
)\}^{\tilde{n}}}\right\vert _{J=\tilde{J}=0}\,, \label{G-gen}%
\end{equation}
is related to the dynamic free energy functional $\Gamma\bigl[\tilde
{s},s\bigr]$ by the Legendre transformation%
\begin{equation}
\Gamma+\mathcal{W}  =\int d^{d}x\,dt\,\bigl[s(\mathbf{x},t)J(\mathbf{x},t)+\tilde{s}(\mathbf{x},t)\tilde{J}(\mathbf{x},t)\bigr] \,,
\label{Leg-Tr}%
\end{equation}
with $\delta\mathcal{W}/\delta J = s$, $\delta\mathcal{W}/\delta \tilde{J} = \tilde{s}$, $\delta\Gamma/\delta s = J$ and $\delta\Gamma/\delta \tilde{s} = \tilde{J}$. $\Gamma$ is of great importance in diagrammatic perturbation theory because it is the generating functional of the irreducible vertex functions. The dependence of $\Gamma$ upon the coupling constant $g$ can be written in
the form
\begin{equation}
\Gamma\bigl[\tilde{s},s;g\bigr]=g^{-2}\Phi\bigl[g\tilde{s},gs;u\bigr]\ .
\label{Phi}%
\end{equation}
The expansion of the functional $\Phi\lbrack\tilde{\varphi},\varphi;u]$ into a
series of $u=G_{\varepsilon}\mu^{-\varepsilon}g^{2}$ yields the
loop-expansion. The zeroth-order term $g^{-2}\Phi\lbrack g\tilde{s},gs;0]$ is
nothing else then the response functional $\mathcal{J}$ (\ref{Funkt}) itself.
Hence, $\mathcal{J}$ constitutes the mean field part of the dynamic free
energy. From the RGE (\ref{RGG}) for the Green's functions it follows that the
RGE for the renormalized dynamic free energy is given by
\begin{align}
&\Big[\mathcal{D}_{\mu}-\frac{\gamma}{2}\int d^{d}x\,dt\,\Big(s(\mathbf{x}%
,t)\frac{\delta}{\delta s(\mathbf{x},t)}+\tilde{s}(\mathbf{x},t)\frac{\delta
}{\delta\tilde{s}(\mathbf{x},t)}\Big)\Big]
\nonumber \\
& \times \Gamma\bigl[\tilde{s},s;\tau ,u;\lambda,\mu\bigr]=0\,. \label{RG-Gamma}%
\end{align}
Exploiting the findings of Sec.~\ref{genFormLog}, the solution the RGE~(\ref{RG-Gamma}) is found to be 
\begin{align}
&\Gamma\bigl[\tilde{s},s;\tau,u;\lambda,\mu\bigr]=
\\
& \Gamma\bigl[Z(w)^{-1/2}%
\tilde{s},Z(w)^{-1/2}s;Z_{\tau}(w)\tau,w;Z_{\lambda}(w)\lambda,l\mu\bigr]\,.
\label{Gamma-l}%
\nonumber
\end{align}

Now we revisit the Langevin equation~(\ref{LangEq}). If the simple epidemic process is supplemented by an additional constant particle source $h$, Eq.~(\ref{LangEq}) is modified to
\begin{align}
\lambda^{-1}\partial_{t}n(\mathbf{x},t) & =\nabla^{2}n(\mathbf{x},t)-\tau
n(\mathbf{x},t)-\frac{g}{2}n(\mathbf{x},t)^{2}
\nonumber \\
&+h+\zeta(\mathbf{x},t)\, .
\end{align}
The extra term in the Langevin equation induces an extra term in the response functional~(\ref{Funkt}),
\begin{equation}
\mathcal{J}\rightarrow\mathcal{J}_{h}=\mathcal{J}-\int d^{d}xdt\,\lambda \, 
h \, \tilde{s}(\mathbf{x},t)\, .
\end{equation}
Hence, the source $\tilde{J}$ is shifted by $\lambda h$, and, to obtain the true generating functional of the irreducible vertex functions, we have to translate the field $s$ by its mean value $M$. Then it
is not difficult to see~\cite{JaKuOe99} that 
\begin{align}
\lambda h&=\left.  \frac{\delta\Gamma}{\delta\tilde{s}(\mathbf{x},t)}\right\vert _{\tilde{s}=0,s=M}=\Gamma_{1,0}(M,\tau)
\nonumber \\
& =:g^{-1}\Phi_{1,0}(gM,\tau,u) \label{Quelle}%
\end{align}
constitutes the EQS that relates the particle source
to a constant mean particle density $M$ for a given $\tau$. By simple
dimensional considerations we find that $\Phi_{1,0}$ obeys the scaling
relation%
\begin{align}
\Phi_{1,0}\bigl(gM,\tau,u;\lambda,\mu\bigr)  &  =\lambda\mu^{4}\Phi
_{1,0}\bigl(gM/\mu^{2},\tau/\mu^{2},u;1,1\bigr)\nonumber\\
&  =:\lambda\mu^{4}F\bigl(gM/\mu^{2},\tau/\mu^{2},u\bigr)\,.
\label{Phi-dimlos}%
\end{align}
A one-loop calculation \cite{JaKuOe99} yields%
\begin{align}
F(x,y,u)/x&=\bigl(y+x/2\bigr)+u\frac{(x+y)}{2}\Big[\ln(x+y)-1\Big]
\nonumber \\
&+O(u^{2})
\label{ZGl-1L}%
\end{align}
Exploiting now Eqs.~(\ref{Gamma-l}) and (\ref{Phi-dimlos}) we get the result
\begin{align}
 \label{gen-EQS}
h  &  =\sqrt{G_{\varepsilon}/w}(l\mu)^{2+d/2}Z_{\lambda}(w)Z(w)^{-1/2}
\\
&  \times F\bigl(\sqrt{w/G_{\varepsilon}}Z(w)^{-1/2}M/(l\mu)^{d/2},Z_{\tau
}(w)\tau/(l\mu)^{2},w\bigr)\,,\nonumber
\end{align}
for the general scaling form of the EQS. Here, $l=l(w)$ is given
by Eqs.~(\ref{Char-u}) and (\ref{l(w)}).

\subsection{Behavior at the upper critical dimension $d=4$}

In four dimensions we have $G_{\varepsilon=0}=1/(4\pi)^{2}$. Following our
work for $d<4$ in \cite{JaKuOe99} we will cast the EQS in a
parametric form. To this end we make the ansatz
\begin{equation}
\tau/\mu^{2}=R\, (1-\theta)\,,\qquad4\pi M/\mu^{2}=f_{M}(w) \, R\, \theta\label{Ansatz}%
\end{equation}
so that $R=0$ corresponds to the critical point. The parameter $\theta$ describes the
crossover from the absorbing to the active phase. The source is zero for
$\theta=0$ and $\theta=\theta_{0}>1$. We expect that $\theta_{0}=2$. After inserting
the ansatz~(\ref{Ansatz}) into the scaling form of the EQS~(\ref{gen-EQS}) we choose the parameter $w$ so that
\begin{equation}
l^{-2}Z_{\tau}(w)R=c\label{c-Wahl} \, ,
\end{equation}
where $c$ is a convenient dimensionless positive constant. Defining%
\begin{equation}
p(w)=Z_{\tau}(w)^{-1}Z(w)^{-1/2}w^{1/2}f_{M}(w)\label{p(w)}%
\end{equation}
we arrive at%
\begin{align}
4\pi h/\mu^{2}&=R^{2}Z_{\lambda}(w)Z_{\tau}(w)^{2}(wZ(w))^{-1/2}
\nonumber \\
& \times c^{-2} \, F(cp(w)\theta,c(1-\theta)),w)\,.\label{h-det}%
\end{align}
Next we determine the function $p(w)$ so that $F(cp(w)\theta,c(1-\theta)),w)$ is analytic in $\theta$ and has an expansion in $w$. Using the 1-loop result~(\ref{ZGl-1L}) we obtain readily $\theta_{0}=2+O(w^{2})$ and
\begin{equation}
p(w)=1+w(1-\ln c)/4+O(w^{2})\,.\label{p(w)-res}%
\end{equation}
It follows that
\begin{equation}
2 \, c^{-2} \, F(cp(w)\theta,c(1-\theta)),w)=\theta(2-\theta)+O(w^{2})\,.\label{F-res}%
\end{equation}

Using Eqs.~(\ref{l(w)}), (\ref{Q(w)}) and (\ref{c-Wahl}) in conjunction with a rescaling of the arbitrary dimensionless variable $R$, we get
\begin{equation}
r:=-\frac{3}{4}\ln R=w^{-1}-a_{1}\ln w+O(w)\,. \label{r(w)}%
\end{equation}
The constant $a_{1}$ is given by
\begin{equation}
a_{1}=-\frac{\beta_{2}\kappa_{1}+2\beta_{3}}{2\beta_{2}}=\frac{133}{192}%
+\frac{53}{96}\ln\frac{4}{3}=0.851533\,. \label{a_1}%
\end{equation}
Exploiting Eq.~(\ref{t(w)}) we obtain for the dimensionless coupling constant as a
function of $r$ the asymptotic expression
\begin{equation}
w=r^{-1}\exp\bigg[a_{1}\frac{\ln r}{r}+O\Big(\frac{\ln^{2}r}{r^{2}},\frac{\ln
r}{r^{2}},\frac{1}{r^{2}}\Big)\bigg]\,. \label{w(r)}%
\end{equation}
Collecting our results we get finally the equation of state in parametric
form
\begin{subequations}
\label{Zust.Gl.}%
\begin{align}
\tau/\tau_{0}  &  =R(1-\theta)\,,\label{tau}\\
M/M_{0}  &  =R\theta\bigl(w^{-1}+\mathcal{Y}\bigr)^{1/3}\exp\bigl(-c_{M}%
w+O(w^{2})\bigr)\nonumber\\
&  =R\theta\bigl(r+\mathcal{Y}\bigr)^{1/3}\bigg[1-\frac{b_{M}\ln r+c_{M}}%
{r}\nonumber\\
&  +O\Big(\frac{\ln^{2}r}{r^{2}},\frac{\ln r}{r^{2}},\frac{1}{r^{2}%
}\Big)\bigg]\,,\label{M}\\
h/h_{0}  &  =R^{2}\theta(2-\theta)\exp\bigl(-w/6+O(w^{2})\bigr)\nonumber\\
&  =R^{2}\theta(2-\theta)\bigg[1-\frac{1}{6r}+O\Big(\frac{\ln r}{r^{2}%
}\Big)\bigg]\,. \label{h}%
\end{align}
The constants $b_{M}$, $c_{M}$, and $c_{h}$ are given by
\end{subequations}
\begin{subequations}
\begin{align}
b_{M}  &  =\frac{a_{1}}{3}=\frac{133}{576}+\frac{53}{288}\ln\frac{4}%
{3}=0.283844,\\
c_{M}  &  =\frac{\kappa_{2}}{\beta_{2}}-\beta_{3}\frac{\kappa_{1}}{\beta
_{2}^{2}} =\frac{71}{768}-\frac{17}{384}\ln\frac{4}{3} 
\nonumber\\
&  =0.079712\,.
\end{align}
Equations~(\ref{w(r)}) and (\ref{M}) show directly that the arbitrary constant
$\mathcal{Y}$ can be eliminated by a rescaling of the nonuniversal constants
$\tau_{0}$, $M_{0}$, and $h_{0}$. However, we keep $\mathcal{Y}$ in our
formulas for the same reasons for which we kept the non-universal constant $\mathcal{Z}$ in Sec.~\ref{logCorr} .

To the order we are working here it is possible to eliminate the parameter $R$ completely. Exploiting Eqs.~(\ref{r(w)}), (\ref{tau}) and (\ref{h}) we can express $R$ in terms of $\tau$ and $h$ as
\end{subequations}
\begin{equation}
\label{Rrel}
R=\bigl(1+O(w)\bigr)\sqrt{\bigl(\tau/\tau_{0}\bigr)^{2}+h/h_{0}}\,.
\end{equation}
Using Eq.~(\ref{Rrel}) we can recast our results stated in Eqs.~(\ref{w(r)}) and (\ref{Zust.Gl.}) as
\begin{equation}
\label{recastRes1}
\sqrt{\bigl(\tau/\tau_{0}\bigr)^{2}+h/h_{0}}=w^{4a_{1}/3}\exp\bigg[-\frac
{4}{3w}+O(w)\bigg]\,,
\end{equation}
and 
\begin{subequations}
\label{recastRes2}
\begin{align}
\frac{\tau/\tau_{0}}{M/M_{0}} &  =\frac{1-\theta}{\theta}\bigl(w^{-1}%
+\mathcal{Y}\bigr)^{-1/3}\bigl[1+c_{M}w+O(w^{2})\bigr]\,,\\
\frac{h/h_{0}}{(M/M_{0})^{2}} &  =\frac{2-\theta}{\theta}\bigl(w^{-1}%
+\mathcal{Y}\bigr)^{-2/3}\bigl[ 1+(c_{M}-1/6)w
\nonumber \\
&+O(w^{2})\bigr]\,.
\end{align}
\end{subequations}

Finally, we recast our results in yet another form. As we will see shortly, this form highlights an interesting and probably important intricacy.  If we take only the leading (1-loop) terms in Eqs.~(\ref{recastRes1}) and (\ref{recastRes2}) into account, our results boil down to the EQS
\begin{equation}
\frac{M}{M_{0}^{\prime}}=\left[ \sqrt{\left(\frac{\tau}{\tau_0} \right)^{2}+ \frac{h}{h_0}
}-\frac{\tau}{\tau_0} \right] \, \left\vert \ln \sqrt{ \left( \frac{\tau}{\tau_0}\right)^{2}+\frac{h}{h_0} }\right\vert^{1/3}
\end{equation}
that is appealingly simple in stature. $M_{0}^{\prime}=(3/4)^{1/3}M_{0}$ is another non-universal constant. Now we switch to the scaled variables
\begin{equation}
X=\frac{\tau/\tau_{0}}{\sqrt{h/h_{0}}}\,,\quad Y=\frac{M/M_{0}^{\prime}}%
{\sqrt{h/h_{0}}\, \left\vert \ln(h/h_{0})\right\vert ^{1/3}}\,. \label{scalvar}%
\end{equation}
By incorporating the corrections coming from the 2-loop order  we obtain
\begin{align}
Y  & =\Big(\sqrt{1+X^{2}}-X\Big)\, \left[ 1+\frac{\ln(1+X^{2})}{\ln(h/h_{0}
)}\right]^{1/3}\nonumber\\
& \times\bigg\{ 1-\frac{1}{12r}\left[ 4a_{1}\ln r+12c_{M}-1-\frac{X}%
{\sqrt{1+X^{2}}}\right]
\nonumber \\
&+O\left(\frac{\ln^{2}r}{r^{2}} \right)\bigg\}\,.\label{asyst}%
\end{align}
In this formula the variable $r$ is given by%
\begin{equation}
r=\frac{3}{8}\left\vert \ln(h/h_{0})\right\vert \, \left[ 1+\frac{\ln
(1+X^{2})}{\ln(h/h_{0})}\right]\,.
\end{equation}
Note that this asymptotic EQS~(\ref{asyst}) is not only a
relation between the scaling variables~(\ref{scalvar}) but that it also includes a
dimensionful non-universal constant $h_{0}$. Only in the limit $\left\vert \ln(h/h_{0})\right\vert
\rightarrow\infty$ we obtain the mean field equation of state although with
the logarithmically corrected scaling variables~(\ref{scalvar}). This intricacy may in deed be
 relevant for the explanation of simulation results~\cite{Lue03}.

\section{Concluding remarks}
\label{concludingRemarks}

In summary, we have investigated logarithmic corrections to scaling in DP by using renormalized dynamical field theory. We calculated the leading and the next to leading logarithmic correction for for the number $N(t)$ of active sites at time $t$ generated by a seed at the origin, the radius of gyration $R(t)$ of the corresponding cluster as well as its survival probability $P(t)$. Moreover we determined the logarithmic corrections to the mean-field equation of state that describes the dependence of the stationary particle density $M(\tau, h)$ upon $\tau$ and an auxiliary external homogeneous source $h$.

Our result involve 2 non-universal scales. The dynamic observables depend on the non-universal time-scale $t_{0}$ and our asymptotic expansions are valid for  times $t \gg t_{0}$. Note that $t_{0}$ may serve as a measure of quality for microscopic models of DP with respect to their suitability for numerical simulations. The smaller $t_0$ is for a given model, the less computer time will be required in order to get good statistics on the critical behavior of DP. Our results for the EQS define the non-universal scale $h_0$ associated with the auxiliary source. The EQS depends on $h_0$ unless the limit $\left\vert \ln(h/h_{0})\right\vert \rightarrow\infty$ is reached. Only in this limit one can expect logarithmically corrected mean-field behavior of the EQS. This should and probably has to be taken into account when numerical data on the EQS is analyzed. The existence of the non-universal scales $t_0$ and $h_0$ can be regarded as 2 examples for Coleman's concept of dimensional transmutation in naively scale-independent field theories. In other words, $t_0$ and $h_0$ are akin to the hadronization scale of quantum chromodynamics.

From the experience one has with other systems at their respective upper critical dimension, in particular linear polymers in $d=4$, we expect that logarithmic corrections are of clear significance with respect to numerical simulations of DP in $d=4$. This expectation is corroborated by recent
simulations~\cite{Lue03}. The aforementioned experience and the fact that we went up to the second logarithmic correction make us confident that our results compare will well with simulations, perhaps even quantitatively. We hope that our analytical estimates trigger an increase effort to determine logarithmic correction for DP numerically with high accuracy.

\begin{acknowledgments}
This work has been supported by the Deutsche Forschungsgemeinschaft via the Sonderforschungsbereich~237 \textquotedblleft Unordnung und gro{\ss }e Fluktuationen\textquotedblright\ and the Emmy Noether-Programm. We thank Sven L\"{u}beck for sharing his simulation results with us before publication. 
\end{acknowledgments}

\appendix

\section{Calculation of Green's functions}

In this Appendix we outline our 1-loop calculations of scaling functions belonging to the Green's functions $G_{1,1}$ and $G_{0,1}$. These calculations provide us with the amplitudes $A_{N}(X_{0})$, $A_{R}(X_{0})$, and $A_{P}(X_{0})$ that enter the second logarithmic corrections for the dynamic observables.

\subsection{The Green's function $G_{1,1}$}
\label{app:G11}
Here we provide some of the details of the calculation that leads from the Dyson equation~(\ref{Dys-Gl}) to our result for $G_{1,1}$ stated in Eq.~(\ref{A_N}). First, we carry out the momentum integration in the self-energy~(\ref{Selbst}). This step leads to
\begin{equation}
\Sigma(\mathbf{q},t)=-\frac{(\lambda g)^{2}\exp\bigl(-\lambda(2\tau
+q^{2}/2)\bigr)}{2(8\pi\lambda t)^{d/2}}\,. \label{Selbst1}%
\end{equation}
Next, we substitute (\ref{Selbst1}) into (\ref{Dys-Gl}). After an integration we obtain 
\begin{align}
G_{1,1}(\mathbf{q},t)  &  =G(\mathbf{q},t)\biggl[1-\frac{u\bigl(2\lambda
\mu^{2}t\bigr)^{\varepsilon/2}}{8\Gamma(1+\varepsilon/2)}\nonumber\\
&  \times\int_{0}^{1}dx\,(1-x)x^{-d/2}\exp\bigl(-\alpha x\bigr)\biggr]\,.
\label{G1_1}%
\end{align}
Here, we introduced the shorthand notation
\begin{equation}
\alpha=\Big(\tau-\frac{q^{2}}{2}\Big)\lambda t. \label{alpha}%
\end{equation}
The remaining integral is calculated in dimensional regularization. A subsequent
$\varepsilon$-expansion yields
\begin{align}
\label{zwieback}
&  \int_{0}^{1}dx\,(1-x)x^{-d/2}\exp\bigl(-\alpha x\bigr)\nonumber\\
&  =(1+\alpha)\Big[-\frac{2}{\varepsilon}-1+\operatorname{Ei}\!\mathrm{n}%
(\alpha)\Big]+1-\exp(-\alpha)+O(\varepsilon)\,,
\end{align}
where the entire exponential integral is given by
\begin{equation}
\operatorname{Ei}\!\mathrm{n}(x)=\int_{0}^{x}dy\,\frac{1-\exp(-y)}{y}
=-\sum_{k=1}^{\infty}\frac{(-x)^{k}}{k!\, k}\,.
\end{equation}

The next step is to remove the $\varepsilon$ poles by employing the
renormalization scheme~(\ref{RenSch}). We let $G_{1,1}\rightarrow\mathring
{G}_{1,1}$, $\lambda\rightarrow\mathring{\lambda}$, $\tau\rightarrow
\mathring{\tau}$, and use the 1-loop results
\begin{subequations}
\label{Z-Fakt}%
\begin{align}
Z=1+\frac{u}{4\varepsilon}+\cdots\,,  &  \quad Z_{\lambda}=1+\frac
{u}{8\varepsilon}+\cdots\,,\\
Z_{\tau}=1+\frac{u}{2\varepsilon}+\cdots\,,  &  \quad Z_{u}=1+\frac
{2u}{\varepsilon}+\cdots\,.
\end{align}
\end{subequations}
Expressing the bare quantities in terms of these renormalization factors and their renormalized counterparts,
\begin{subequations}
\begin{align}
\mathring{G}_{1,1} & = Z G_{1,1} = \Big(1+ \frac{u}{4\varepsilon}\Big) G_{1,1} \, ,
\\
\mathring{\lambda}  &  =Z^{-1}Z_{\lambda}\lambda=\Big(1-\frac{u}{8\varepsilon
}\Big)\lambda\,,
\\
\mathring{\lambda}\mathring{\tau}  &  =Z^{-1}Z_{\tau}\lambda\tau
=\Big(1+\frac{u}{4\varepsilon}\Big)\lambda\tau\,,
\end{align}
\end{subequations}
and using the intermediate results in Eq.~(\ref{G1_1}) combined with Eq.~(\ref{zwieback}) we obtain the renormalized Green's function
\begin{align}
G_{1,1}(\mathbf{q},t)  &  =G(\mathbf{q},t)\biggl\{1+\frac{u}{8}\Big[(1+\alpha
)\Big(\ln(2\lambda\mu^{2}t)+C_{E}\nonumber\\
&  \qquad\qquad\qquad-\operatorname{Ei}\!\mathrm{n}(\alpha)\Big)+\alpha
+\exp(-\alpha)\Big]\biggr\}\,. \label{G1_1(ende)}%
\end{align}
Note that the $\varepsilon$ poles are indeed removed by our renormalization.

Two results important for the logarithmic correction can be extracted from
(\ref{G1_1(ende)}). Upon setting $\alpha=0$ we find
\begin{align}
&  G_{1,1}(\mathbf{q}=0,\lambda\mu^{2}t=X_{0};\tau=0,w;1,1)\nonumber\\
&  =\,1+\frac{1}{8}\Big(\mathcal{Z}+1\Big)w\,, \label{G1_1(0,0)}%
\end{align}
and hence the amplitude $A_{N}(X_{0})$ as stated in Eq.~(\ref{A_N}). Moreover,
we get%
\begin{align}
&  -X_{0}^{-1}\left.  \frac{\partial}{\partial q^{2}}\ln G_{1,1}%
(\mathbf{q},\lambda\mu^{2}t=X_{0};\tau=0,w;1,1)\right\vert _{q^{2}%
=0}\nonumber\\
&  =\,1+\frac{1}{16}\Big(\mathcal{Z}-1\Big)w\,, \label{G1_1(q,0)}%
\end{align}
which leads to our result for $A_{R}(X_{0})$ given in Eq.~(\ref{A_R(X)}).

\subsection{The Green's function $G_{0,1}$}
\label{app:G01}

Here we provide selected details on our 1-loop calculation of $G_{0,1}$ as required in (\ref{newSurvProb}). We start by solving the differential equations~(\ref{Prop-k}) and (\ref{Dgl-shift}). The initial and terminal conditions for the fields necessitate the ansatz
\begin{equation}
\tilde{M}(t)=-\theta(-t)\, K(-t)^{-1}\, . 
\end{equation}
The type of the source term,
$k(t)=k\delta(t)$ with $k\rightarrow\infty$, demands the initial condition
$K(0)=0$. With this information, the differential equation~(\ref{Dgl-shift})
can be transformed without much effort into the integral equation
\begin{equation}
K(t)+\frac{g}{2\tau}=\mathrm{e}^{\lambda\tau t}\biggl(\int_{0}^{t}dt^{\prime
}\,\mathrm{e}^{-\lambda\tau t^{\prime}}K(t^{\prime})^{2}\tilde{T}(-t^{\prime
})+\frac{g}{2\tau}\biggr)\,. \label{K-Int}
\end{equation}
At mean field level, the solution to Eq.~(\ref{K-Int}) is given by Eq.~(\ref{K_0}). Inserting the corresponding $\tilde{M}_{0}(t)=-K_{0}(-t)^{-1}$ into the
differential equation~ (\ref{Prop-k}) we find the modified Gaussian
propagator as stated in Eq.~(\ref{Gmod_0}).

Now to the computation of the diagram depicted in Fig.~\ref{tadpole}b. After some intermediate steps
we obtain
\begin{align}
&  K(t)^{2}T(-t)=\frac{\lambda^{2}g^{2}}{2}K_{0}(t)^{-2}\int_{0}^{t}%
dt^{\prime}\nonumber\\
&  \times\,\frac{K_{0}(t^{\prime})^{3}\exp[2\lambda\tau(t-t^{\prime})]}%
{[8\pi\lambda(t-t^{\prime})]^{d/2}}\,. \label{T_1}%
\end{align}
The further evaluation of Eq.~(\ref{T_1}) is fairly straightforward for
$\tau=0$. After $\varepsilon$ expansion we find
\begin{equation}
K(t)^{2}\tilde{T}(-t)=-\frac{\lambda g^{3}(2\lambda t)^{\varepsilon/2}%
}{16(4\pi)^{d/2}}\Big(-\frac{6}{\varepsilon}+\frac{3}{2}\Big)\,.
\end{equation}
Insertion of this intermediate result into (\ref{K-Int}) yields
\begin{equation}
K(t)=\frac{g\lambda t}{2}\biggl[1+\frac{u(2\lambda\mu^{2}t)^{\varepsilon/2}%
}{4\Gamma(1+\varepsilon/2)}\Big(-\frac{3}{\varepsilon}+\frac{9}{4}%
\Big)\biggr]\,. \label{K-Ergeb}%
\end{equation}
Next, we renormalize. Indicating the consistency of our previous steps, the
appropriate combination of renormalization factors $Z_{u}{}^{1/2} Z^{-1} =1+3u/(4\varepsilon)+\cdots$ cancels the $\varepsilon$ pole in
(\ref{K-Ergeb}). The renormalized $K(t)$ reads
\begin{equation}
K(t)=\frac{g\lambda t}{2}\biggl[1-\frac{3u}{8}\Big(\ln(\lambda\mu^{2}%
t)+C_{E}-\frac{3}{2}\Big)\biggr]\,.
\end{equation}
Exploiting $G_{0,1}(-t)=K(t)^{-1}$ and $\lambda\mu^{2}t=X_{0}$ as well as
recalling the definition of $\mathcal{Z}$ we finally obtain
\begin{align}
&  2\pi X_{0}G_{0,1}(-\lambda\mu^{2}t=X_{0};\tau=0,w;1,1)\nonumber\\
\,  &  \qquad\qquad=w^{-1/2}\biggl[1+\frac{3}{8}\Big(\mathcal{Z}-\frac{3}%
{2}\Big)w\biggr]\,. \label{G0_1(fin)}%
\end{align}

\end{document}